\documentclass[twocolumn]{aastex701}
\usepackage[normalem]{ulem}
\usepackage{amsmath}
\usepackage{graphicx}
\usepackage{tabularx, booktabs}
\newcommand{\add}[1]{\textbf{#1}}
\usepackage{xcolor}
\definecolor{cutcolor}{rgb}{1,0,0}      
\definecolor{editcolor}{rgb}{0,0,1}     
\definecolor{commentcolor}{rgb}{0,0.5,0} 

\renewcommand{\add}[1]{#1}

\shorttitle{Bridging Theory and Observation: Synthetic FIR Insights into Star Formation Efficiency}
\shortauthors{Escamilla et al.}

\begin{document}

\title{Bridging Theory and Observation: Synthetic FIR Insights into Star Formation Efficiency}

\correspondingauthor{Alexander Escamilla}
\email{alex.escamilla2244@gmail.com}

\author{Alexander Escamilla}
\affiliation{Department of Astronomy, San Diego State University, San Diego, CA 92182, USA}
\email{alex.escamilla2244@gmail.com} 

\author{Michael Y. Grudi\'{c}}
\affiliation{Center for Computational Astrophysics, Flatiron Institute, New York, NY 10010, USA}
\email{mgrudic@flatironinstitute.org} 

\author[orcid=0000-0003-4423-0660]{Anna L. Rosen}
\affiliation{Department of Astronomy, San Diego State University, San Diego, CA 92182, USA} 
\affiliation{Computational Science Research Center, San Diego State University, San Diego, CA 92182, USA}
\email{alrosen@sdsu.edu}

\begin{abstract}
The star formation efficiency per free-fall time ($\epsilon_{\rm ff}$) quantifies how efficiently giant molecular clouds (GMCs) convert gas into stars, and is often observed to be orders of magnitude lower than expected for free-fall collapse. Observers typically estimate $\epsilon_{\rm ff}$ by mapping FIR dust emission to infer gas surface densities ($\Sigma_{\rm gas}$) and counting embedded protostars, assuming simplified lifetimes and masses. Using the fiducial \textsc{starforge} radiation--magnetohydrodynamics simulation of a $M_{\rm cl} = 2\times10^4\, \rm M_\odot$ GMC that self-consistently models star cluster formation and stellar feedback, we generate synthetic far-infrared (FIR) observations and apply the same methodologies used in GMC surveys to investigate this discrepancy. We present the first $\epsilon_{\rm ff} -\Sigma_{\rm gas}$ analysis in a fully feedback-regulated GMC simulation that resolves \add{the formation of stellar systems}. Our synthetic measurements reproduce the low observationally inferred efficiencies, showing that feedback-regulated star formation naturally produces $\epsilon_{\rm ff} \sim 1$--$3\%$ without requiring extreme initial conditions. We also find that $\epsilon_{\rm ff}$ varies strongly over a GMC's lifetime, suggesting that much of the observed scatter reflects evolutionary sampling rather than intrinsic cloud-to-cloud differences. \add{Finally, by comparing observational and simulation definitions, we show that methodological assumptions introduce systematic biases, with a transition near $\log \Sigma_{\rm gas} \approx 2.3 \rm\,  M_\odot{\rm pc}^{-2}$ that depends on resolution. Above this threshold, smoothing of dense structure increases both the inferred free-fall time and enclosed gas, with the latter dominating and suppressing $\epsilon_{\rm ff}$. Below the threshold the discrepancies are primarily driven by star formation rate (SFR) assumptions.}
\end{abstract}

\vspace*{-0.3\baselineskip}
\keywords{Star formation (1569); Star forming regions (1565);Giant molecular clouds (653); Protostars (1302); Young stellar objects (1834); Early stellar evolution (434); Far infrared astronomy (529); Infrared astronomy (786);
Dust continuum emission (412); Magneto-hydrodynamical simulations (1966)}
\section{Introduction}
\label{sec:Introduction}

Star formation is a fundamental process that shapes the structure and evolution of galaxies, sets the properties of star clusters, and the initial conditions for planet formation \citep{McKee2007}. Despite its importance, the physical mechanisms regulating the conversion of cold, dense interstellar gas into stars remain elusive due to the complex interplay of multiple physical processes, including magnetic fields, turbulence, and stellar feedback that act within giant molecular clouds (GMCs) before they ultimately collapse into protostars \citep{bergin2007GMC, Padoan2014}. These mechanisms operate across many scales, making star formation a complex, non-linear physical process.

One of the most challenging aspects of studying star formation is the extended lifetimes of GMCs ($\geq 10 \,\mathrm{Myr}$), which makes it difficult to obtain a complete picture of star formation from observations alone. Thus, a convenient way to quantify cloud-scale star formation is through the star formation efficiency per free-fall time, $\epsilon_{\rm ff}$, a dimensionless measurement that indicates the fraction of a cloud's mass that is converted to stars during one gravitational free-fall time:

\begin{equation} \label{eq:epsff}
\epsilon_{\rm ff} = \frac{\dot{M}_{\star}}{ M_{\rm gas}/t_{\rm ff}},
\end{equation}
where
$\dot{M}_{\star}$ is the star formation rate (SFR), $M_{\rm gas}$ is the GMC mass, and $t_{\rm ff}$ is the cloud gravitational free-fall time.

\begin{figure*}[!t]
    \centering
    \includegraphics[width=0.55\textwidth]{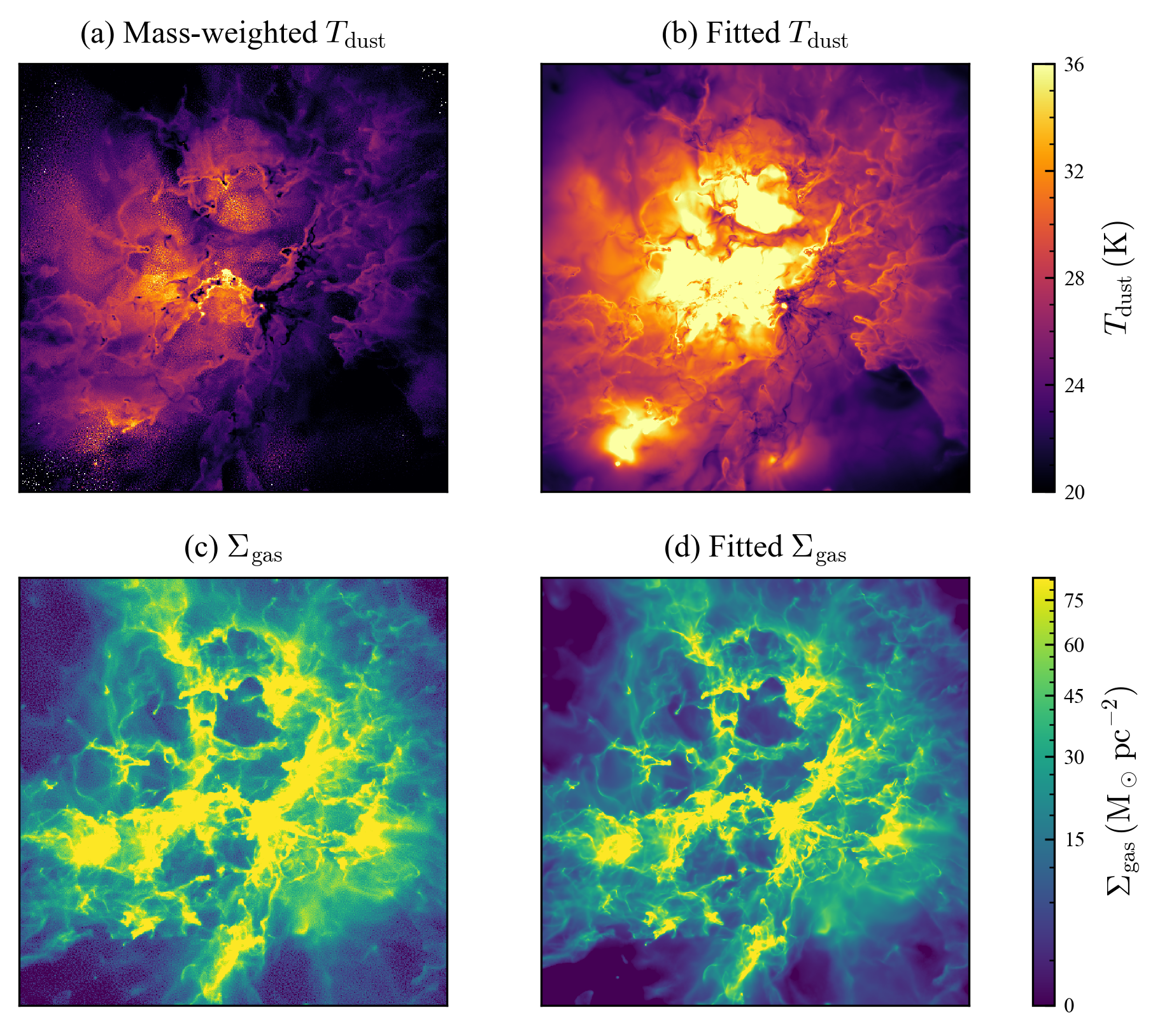}
    \caption{Comparison of intrinsic STARFORGE dust properties with fitted values from our synthetic observations. (a) Mass-weighted dust temperature $T_{\rm dust}$ from the simulation. (b) Fitted $T_{\rm dust}$ recovered from SED modeling. (c) Intrinsic gas surface density $\Sigma_{\rm gas}$. (d) Fitted $\Sigma_{\rm gas}$ derived from $500\,\mu\mathrm{m}$ optical depth assuming a 1\% dust-to-gas ratio. 
    While SED fitting slightly overestimates dust temperatures, it reproduces the cloud morphology and yields reasonable surface density estimates, demonstrating the reliability of our synthetic observation pipeline.}

    \label{fig:four_panel_sigma}
\end{figure*}

$\epsilon_{\rm ff}$ has historically exhibited a persistent mismatch between computational predictions and observational measurements \citep{starproblems2014}. \citet{1974ApJ...192L.149Z} first identified this discrepancy, measuring values of $\epsilon_{\rm ff} \leq 1\%$. In comparison, idealized theoretical models that only accounted for gravity predicted efficiencies of order $100\%$ from clouds \citep{100eff}. These results suggested either observational biases or missing physical mechanisms, such as turbulence \citep{mac2004}, magnetic fields \citep{federrath2012}, and stellar feedback -- the injection of energy and momentum into the ISM by young stars \citep{2015NewAR..68....1D, grudic2019SFE}. Since then, several studies have demonstrated how these mechanisms suppress gravitational collapse and regulate the conversion of interstellar gas into stars \citep{STARFORGE, Anna+Krumholz+2020, Appel+2023}. However, these models must be compared with observations directly to determine which of these physical mechanisms is primarily responsible for inefficient star formation in star-forming clouds.

None of the quantities in Eq.~\eqref{eq:epsff} — the stellar accretion rate, total gas mass, free-fall time, and the gas density — are directly measurable, or even well-defined in realistic systems with substructure and line-of-sight blending. Consequently, observers must infer $\epsilon_{\rm ff}$ through indirect methods, which primarily include: counting young stellar objects (YSOs), assuming average YSO lifetimes and masses, and adopting geometric assumptions about cloud structure to estimate $t_{\rm ff}$ \citep{Evans2009, Dunham2014, Pokhrel2021, Zipeng2022}. Observational studies of star formation depend on far-infrared (FIR) surveys to identify embedded protostars by tracking FIR emission from dust heated by young stars. This approach yields maps of star-forming regions and estimates of gas surface densities. Molecular gas mass is inferred from FIR dust emission by converting the inferred dust column densities and adopting a dust-to-gas ratio \citep{Evans2009, Sadavoy2013, Gutermuth2009}. Each step introduces uncertainties affecting the molecular gas estimates used to calculate GMC masses and free-fall timescales. \add{Additionally, background active galactic nuclei (AGN) contamination can skew YSO counts, leading to inflated SFRs \citep{pokhrel2020, 2023AGNMillstone}}. This complicates the interpretation of whether observed variations in $\epsilon_{\rm ff}$ stem from intrinsic physical differences or result from observational biases and simplified assumptions.

In contrast, high-fidelity numerical simulations provide the ``ground truth'' by capturing the evolution of gas and stars in a GMC, thereby enabling a more comprehensive analysis of $\epsilon_{\rm ff}$ \citep{syntehticobs}. Previous simulation studies have addressed this challenge by recovering $\epsilon_{\rm ff}$ from synthetic observables. Early work by \citet{offner2009} demonstrated that radiative feedback from protostars heats the surrounding gas in star-forming regions and reduces unrealistically high $\epsilon_{\rm ff}$ toward more observationally consistent values, establishing stellar feedback as a regulatory mechanism; however, that study only included radiative feedback, neglected other forms of stellar feedback (e.g., protostellar outflows), and did not capture the formation of high-mass stars. Most recently, \citet{Suin2024} examined individual feedback mechanisms in separate simulations, finding that different feedback mechanisms ultimately reduced theory–observation discrepancies in $\epsilon_\text{ff}$, but did not self-consistently capture individual star formation due to resolution constraints. These limitations highlight the need for a high-resolution simulation suite capable of capturing stellar systems that includes all major feedback processes inherent to star formation.

Quantifying the degree to which observational biases versus intrinsic physical processes drive the observed discrepancies remains an open question. Thus, a direct comparison between observational and simulation approaches is needed to untangle the source of this discrepancy. In this study, we generate synthetic FIR observations from a radiation-magnetohydrodynamics (RMHD) star formation simulation suite, thereby enabling a controlled test to determine how systematic observational biases affect the inferred values of $\epsilon_\text{ff}$. We utilize STARFORGE, a simulation framework implemented in \textsc{GIZMO} \citep{hopkinsgizmo2017}. STARFORGE self-consistently incorporates all major stellar feedback processes while resolving \add{the system initial mass function (IMF) down to $0.1\, \rm M_{\odot}$ on GMC scales \citep{STARFORGE}; we note that while our simulation tracks individual sink particles, the imposed $18 \rm\, AU$ gravitational softening limit and accretion radius prevent the resolution of close binaries and disk fragmentation,  thus these particles may represent either single stars or unresolved multiple systems.}

While \citet{STARFORGE} measured the global $\epsilon_{\rm ff}$, a detailed analysis of the $\epsilon_{\rm ff}-\Sigma_{\rm gas}$ relationship has not yet been performed. We address this gap by producing synthetic FIR observations of the STARFORGE simulation suite and present the first $\epsilon_{\rm ff}-\Sigma_{\rm gas}$ test in a full-feedback GMC by directly comparing our observationally resolved estimates of $\epsilon_\text{ff}$ against the simulations' intrinsic values throughout its evolution.
We follow the observational methodology introduced by \citet{Pokhrel2021}, which found that different surface density thresholds lead to variations in $\epsilon_{\rm ff}$ across star-forming GMCs. Any discrepancies that persist after applying these methods may indicate missing physical processes in the STARFORGE simulations. This approach also aids in determining how factors like spatial resolution and AGN contamination influence measurements of $\epsilon_{\rm ff}$. We then examine variables such as cloud evolutionary stage, line-of-sight resolution, choice of methodology, and projected contamination may affect trends in the observationally-inferred values of $\epsilon_\text{ff}$, and investigate whether such variations arise from the underlying physics and evolution, or are primarily due to observational limitations. 

The structure of this paper is as follows: Section~\ref{sec:methods} describes the numerics and synthetic observation pipeline employed on the simulated snapshots. Section~\ref{sec:results} presents our main findings, Section~\ref{sec:discussion} discusses their implications for interpreting observed star formation efficiencies, and lastly Section~\ref{sec:conclusion} presents our conclusions.

\section{Methods}\label{sec:methods}
\subsection{Simulations}
We conduct our analysis using STARFORGE, a star formation simulation framework built on GIZMO, a magnetohydrodynamics (MHD) code that employs a mesh-free, Lagrangian finite mass method to solve the equations of ideal MHD \citep{hopkins2016, hopkinsgizmo2017} coupled with self-gravity, stellar dynamics, and multi-band radiative transfer. STARFORGE incorporates key stellar feedback processes that includes: stellar irradiation, thermal re-emission by interstellar dust, protostellar outflows, stellar winds, and supernovae \citep{STARFORGE}. Protostars are modeled as accreting sink particles tied to a sub-grid protostellar evolution model following the methods described in \citet{offner2009}. This setup allows a multi-physics approach that captures self-consistent star formation, stellar feedback, and global GMC dynamics.

\begin{figure}[htbp]
    \centering
    \hspace*{-5mm}
    \includegraphics[width=1\columnwidth, trim=08 08 0 0, clip]{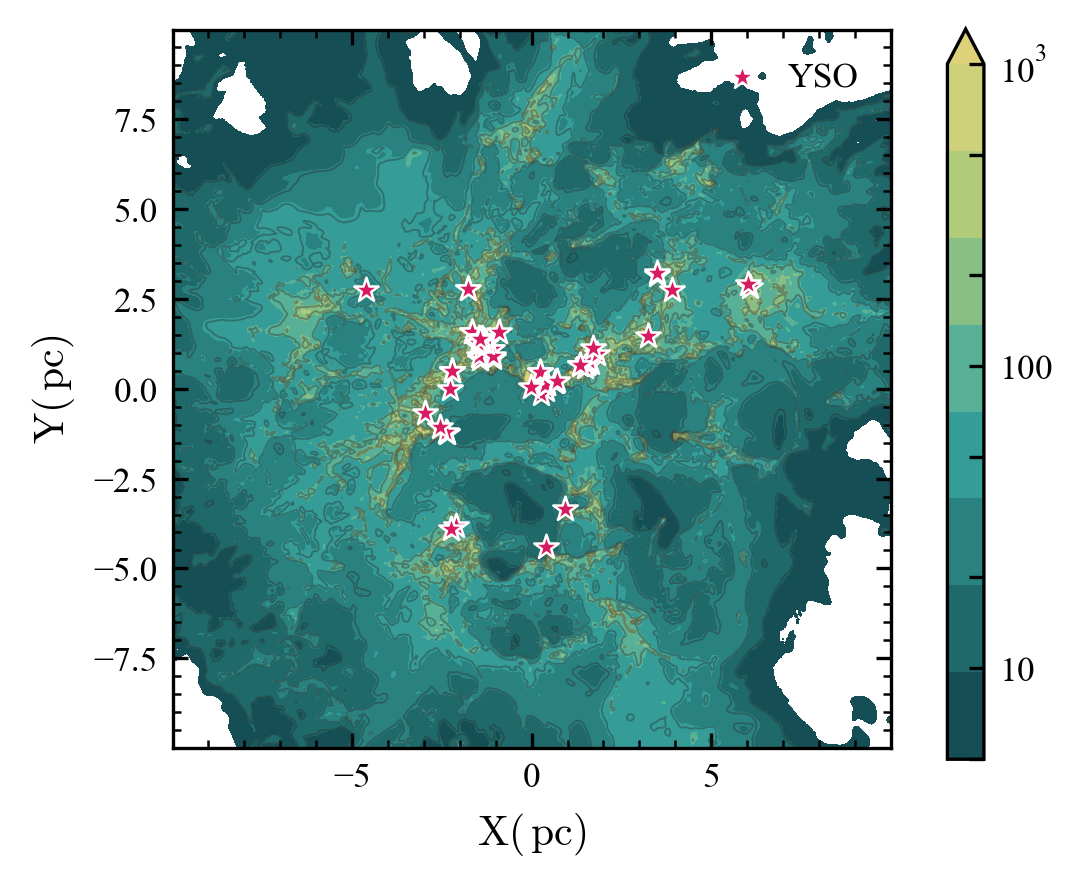}
    \caption{Sample synthetic observation of a snapshot of our fiducial simulation. The color scale represents gas surface density in units of $\rm M_\odot\,\mathrm{pc}^{-2}$. Overlaid contours trace the density structures and show the apertures used to compute $\epsilon_{\rm ff}$. 
Magenta stars denote young stellar objects (YSOs), which predominantly form within the densest regions of the cloud.}
    \label{fig:SyntheticObs}
\end{figure}

We used the fiducial run from the STARFORGE suite designed to model typical Milky Way GMCs. The initial conditions include a turbulent cloud with a mass of $2 \times 10^4 \, \rm M_\odot$ and a radius of $\approx 10\, \rm pc$, giving a mean surface density of $\Sigma_{\rm gas} = M/\pi R^2 = 64\,\rm M_\odot\mathrm{pc}^{-2}$ consistent with measurements from nearby GMCs \citep{Heyer2009, Lada2010}. This simulation reproduces key observational benchmarks, including the stellar system initial mass function \citep[IMF;][]{2010ARA&A..48..339B} and the observed star formation efficiency per free-fall time, $\epsilon_{\rm ff}$. Here we summarize the initial conditions; for a full description, see \citet{STARFORGE}. The GMC is embedded in a low-density ambient medium within a $100\, \rm pc$ simulation box to minimize boundary effects.

An initial random supersonic turbulent velocity field with a power-law spectrum, $P(k) \propto k^{-2}$, is imposed on the cloud and scaled to yield a virial parameter $\alpha_{\text{turb}} \approx 2$ at initialization, where $\alpha_{\text{turb}} = \frac{5 R \sigma^2}{3 G M}$. A uniform magnetic field is imposed to produce an initial mass-to-flux ratio of $\mu \approx 1.3$ \citep{mass2fluxratio}, so the cloud is magnetically supercritical, consistent with observational constraints \citep{2012ARA&A..50...29C}. The simulation has a mass resolution of $10^{-3}~\rm M_\odot$ per gas element. Sink particles, representing protostellar systems, are introduced once all sink formation criteria are satisfied, following \citet{starforgemethods}. Each sink particle evolves according to the subgrid protostellar model calibrated by \citet{offner2009}, which updates the proto-star's mass, radius, and luminosity based on its accretion history. These protostellar properties are inputs to the subgrid stellar feedback models that include radiative feedback, photo-ionization, bipolar outflows, and stellar winds. \citep{starforgemethods}.

As the simulation evolves, the initial gravitational collapse of the GMC leads to the formation of dense cores that collapse to form protostars. These protostars inject momentum and energy into the interstellar medium (ISM) as they accrete mass, altering the cloud’s structure, slowing collapse, and eventually dispersing the cloud \citep{STARFORGE}. The simulation is evolved until it is disrupted by feedback, which occurs at $\approx 10\, \rm Myr$ ($\approx 2.5$ free-fall times).

\subsection{Synthetic Observations and Radiative Transfer}
\label{sec:Synthetic_Obs}

We generate synthetic dust emission maps at FIR wavelengths of 150, 250, 350, and 500 $\mu$m to match observational surveys conducted with the SPIRE instrument aboard the \textit{Herschel} Space Observatory \citep{pokhrel2016}. \add{These maps are produced by post-processing the three-dimensional gas density and dust temperature values from the simulation using \textbf{\texttt{meshoid}}, a Python package for analyzing meshless particle data \citep{Grudic_meshoid}, which integrates thermal dust emission from gas particles along lines of sight, using each particle's smoothing length to distribute its contribution across the resulting synthetic maps.}
Dust emission obeys the time-independent radiative transfer equation
\begin{equation}\label{radiative_equation}
\resizebox{0.9\hsize}{!}{$
\hat{\Omega} \cdot \nabla I_\nu(\mathbf r,\hat{\Omega})
= -\,\kappa_\nu(\mathbf r)\,\rho(\mathbf r)\, I_\nu(\mathbf r,\hat{\Omega})
+ \kappa_\nu(\mathbf r)\,\rho(\mathbf r)\,B_\nu(T)
$}
\end{equation}

\noindent
where $I_\nu$ is the specific intensity, $\kappa_\nu$ is the dust opacity, $\rho$ is the gas density, and $B_\nu(T)$ is the Planck function. \add{Our approach assumes optically thin thermal emission from interstellar dust and models it directly, without Monte Carlo radiative transfer or ray tracing. As a result, our method is computationally efficient while providing sufficient accuracy for our analysis.
Under these assumptions, we use a modified blackbody model to fit the intensity of emission at fixed wavelengths, which are treated as delta functions (i.e. observations at 150, 250, 350, and 500 $\mu\rm m$)} in accordance with traditional observational modeling \citep{pokhrel2016}:

\begin{equation}\label{modifiedSED}
I_\nu = \kappa_0 \left( \frac{\nu}{\nu_0} \right)^{\beta} B_\nu(T)\Sigma_{\rm gas}
\end{equation}
\noindent
where $\Sigma_{\rm gas}$ is the gas surface density, $\beta=1.8$ is the fixed emissivity index, and $\kappa_0$ is the reference opacity from the Astrodust+PAH model \citep{hensley2022}. A dust-to-gas ratio of 1\% is assumed throughout, consistent with observational SED fitting techniques applied to Herschel data \citep{SPIRES2010}. Integrating Eq.~\eqref{modifiedSED} along each line of sight yields the fitted optical depth, which is then converted into mock observed $\Sigma_{\rm gas}$ by dividing by $\kappa_{0}$. We specifically use the fit at $500\, \mu m$ to optimize sensitivity to cold, dense gas. Our synthetic observations achieve a maximum spatial resolution of $0.01\,  \rm pc \approx2000\,  \rm AU$. These maps reproduce the overall cloud morphology and yield gas surface densities that are reasonable when compared with the simulation’s intrinsic values, thereby validating our synthetic observation pipeline (Figure~\ref{fig:four_panel_sigma}).

\begin{figure*}[!t]
    \centering
    \includegraphics[width=0.7\linewidth]{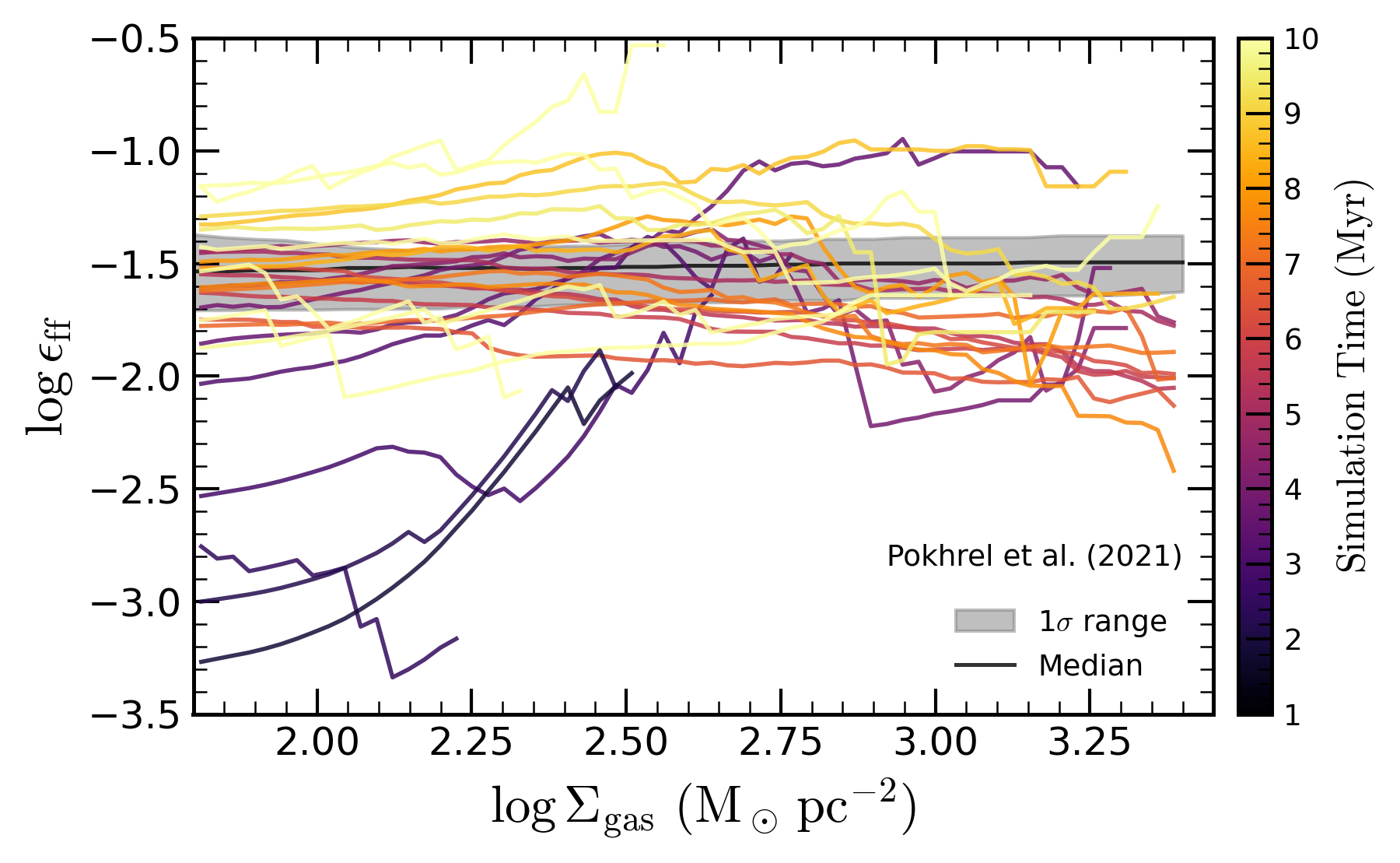}
    \caption{$\epsilon_{\rm ff}$ versus $\Sigma_{\rm gas}$ for the fiducial GMC over $\sim$10\,Myr ($\approx 2.5\,t_{\rm ff}$), where individual tracks represent binned medians per snapshot to highlight physical trends. Colors show the evolutionary stage from early (purple) to late (yellow). Efficiencies start very low and rise sharply in the first $3\text{--}5$\,Myr. By 5\,Myr, the tracks form a brief plateau before exhibiting a decrease at high densities; this declining trend at high $\Sigma_{\rm gas}$ persists through the $6\text{--}7$\,Myr phase (orange/red). In the $8\text{--}10$\,Myr phase (yellow), the relation becomes more stochastic; while most tracks show a stronger decrease at high densities as feedback disperses the cloud, local maximums in efficiency occur during the final stages of cluster formation. The gray band indicates the $1\sigma$ scatter of Milky Way GMCs from \citet{Pokhrel2021}, with a solid line of the median.}
    \label{big_sfe_plot}
\end{figure*}

\subsection{Young Stellar Objects}
\label{subsec:YSO}

To correlate the number of YSOs with SFRs and local gas densities, similar to methods employed in observational studies, \add{we utilize a hybrid approach for identifying Class 0 YSOs. We do not attempt to observationally identify class 0 YSOs in our synthetic FIR maps; instead, we use sink particles that represent protostars and protostellar systems, with each particle characterized by a unique spatial location, mass, and formation time.} Throughout the simulation, we classify protostellar sink particles younger than 0.5 Myr as Class 0 YSOs. Class 0 represents the earliest stage of protostellar evolution, typically ranging from 0.1 to 0.5 Myr \citep{Dunham2014}. This classification aligns with observational protostellar definitions used in YSO catalogs derived from Spitzer observations by \citet{Pokhrel2021}. Protostars younger than this threshold correspond to deeply embedded sources characterized by high envelope densities and substantial FIR emission due to their heavily dust-obscured environments. \add{An important caveat is that while this age classification is motivated by observational studies, the identification in our analysis is assigned using the protostar formation time recorded through the simulation's evolution rather than calculated from the synthetic observations. We discuss the implications and justification of this methodological choice in Section~\ref{theory_obser_discussion}. We assign a local surface density to each Class 0 YSO by measuring the surface density at the nearest pixel using a KDTree algorithm and define surface density contours across the entire map at different thresholds following \citet{Pokhrel2021}. These contours determine which YSOs are included in our $\epsilon_{\rm ff}$ calculations at each $\Sigma_{\rm gas}$ threshold in the cloud.}

To address potential AGN contamination in FIR surveys, we inject synthetic AGN into our YSO catalog at a rate of $9\,\mathrm{deg}^{-2}$ \citep{pokhrel2020}. We place the cloud at a distance of $0.6\,\mathrm{kpc}$, the median of the cloud distances covered in the study by \citet{Pokhrel2021}, which sets the angular size of the field on the sky, and draw the number of AGN from a Poisson distribution with $\lambda = A_{\rm field}\times 9\,\mathrm{deg}^{-2}$, where $A_{\rm field}$ is the projected sky area of the box from our simulation.

\subsection{Measuring Star Formation Efficiency} \label{Star Formation Efficiency Calculation}

We calculate $\epsilon_{\rm ff}$ from our synthetic dust emission maps and YSO catalogs, following the observational methods employed in \citet{Pokhrel2021}. In this approach, the SFR is estimated as
\begin{equation}\label{eq:SFR}
    \dot{M}_{\star} \;=\; \frac{N_{\mathrm{YSOs}}\langle M_{\mathrm{YSO}} \rangle}{\langle t_{\mathrm{YSO}} \rangle}
\end{equation}

where $N_{\mathrm{YSOs}}$ is the number of Class~0 YSOs within the defined contour, and $\langle M_{\mathrm{YSO}} \rangle = 0.5\, \rm  M_\odot$, \add{following previous observational estimates \citep{Dunham2014, Evans2009} and consistent with the peak of the system IMF \citep{IMF, KrouparIMF2002}}. With both $\langle M_{\mathrm{YSO}} \rangle$ and $\langle t_{\mathrm{YSO}} \rangle$ treated as constants, with the SFR scaling directly with $N_{\mathrm{YSOs}}$. The gravitational free-fall time within each contour is then
\begin{equation}\label{spherical_free_fall}
    t_{\mathrm{ff}} = \sqrt{ \frac{3\pi}{32 G \rho} }
\end{equation}
where we assume the gas enclosed by each surface density contour forms a sphere. Following \citet{Krumholz+2012a}, the line-of-sight dimension is taken to be comparable to the projected size of the contour, giving
\begin{equation}
    \rho = \frac{3\sqrt{\pi}\, M_{\mathrm{gas}}}{4\, A^{3/2}}
\end{equation}
where $G$ is the gravitational constant, $M_{\rm gas}$ is the enclosed gas mass, $\rho$ the mean density, and $A$ the projected area of the contour. An example of the synthetic observation of our gas cloud is shown in Figure \ref{fig:SyntheticObs}.

\section{Results}\label{sec:results}

\subsection{Star Formation Efficiency vs. Gas Surface Density}

We present the main result of our work in Figure \ref{big_sfe_plot}, where we replicate observational trends from \citet{Pokhrel2021}. The gray shaded band shows the $1\sigma$ scatter of their GMC sample and a solid line of their median against our fiducial GMC values. We estimate Equation \eqref{eq:epsff} using \eqref{eq:SFR} and \eqref{spherical_free_fall}, turning it into an observational methodology at multiple surface density thresholds ($\Sigma_\text{gas} \sim 100 - 2500\, \rm M_{\odot}\, \mathrm{pc^{-2}}$) for several snapshots taken from the full simulation timeline ($\approx 10$ Myr). To highlight the physical trajectories across these snapshots and mitigate stochastic numerical noise, each track represents the median $\epsilon_{\rm ff}$ calculated within log-spaced density bins. From our synthetic observations, this figure demonstrates our inferred values of $\epsilon_\text{ff}$. We find that $\epsilon_{\rm ff}$ varies both spatially and temporally with $\Sigma_{\rm gas}$ across the GMC with values ranging from $0.05\%$ to $10\%$, with the majority of our estimates occurring between $1-3\%$ in agreement with observationally measured values.

\begin{figure}[hpbt] 
    \hspace*{-1.1cm}
    \centering\includegraphics[width=0.4\textwidth, trim=07 08 0 0, clip]{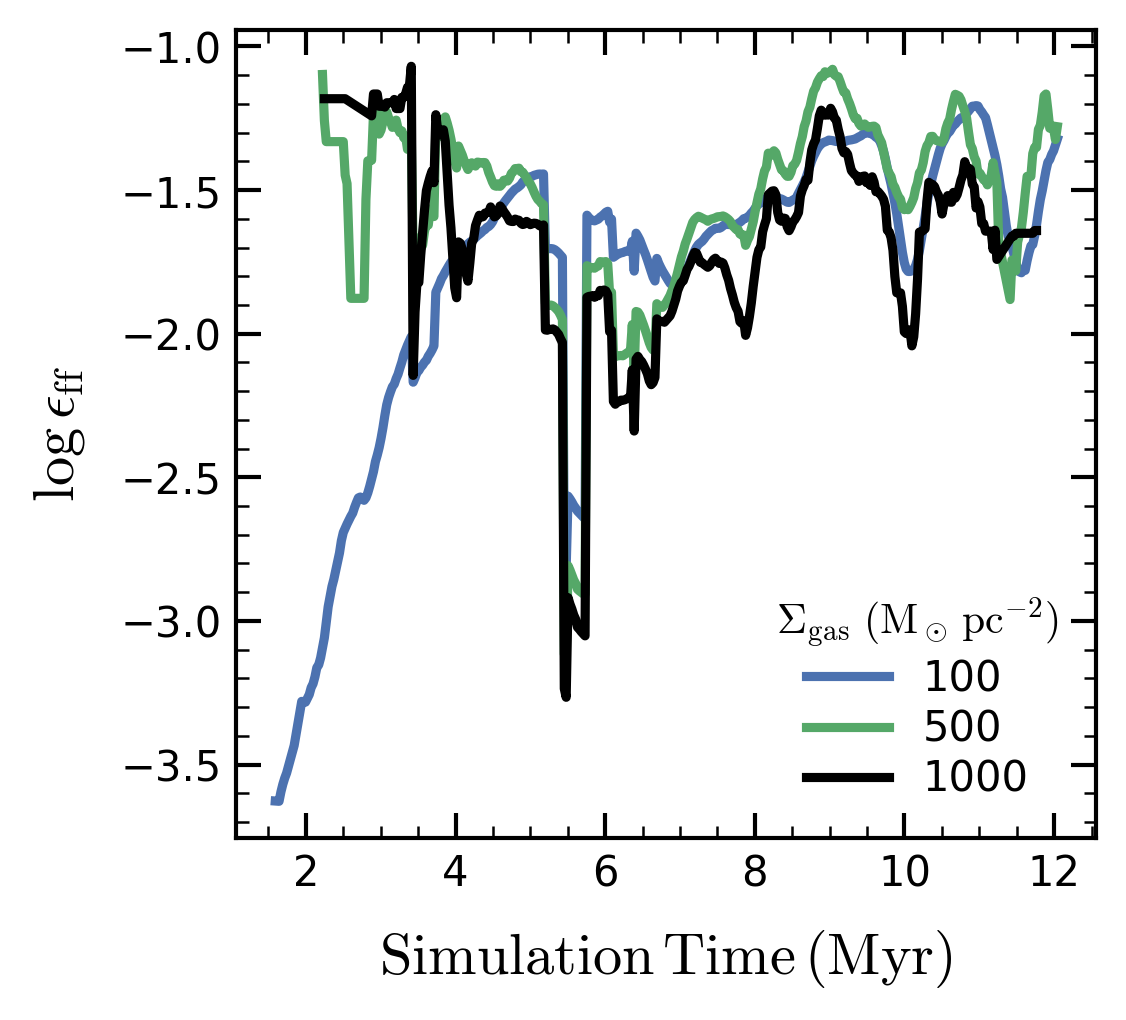}
    \caption{$\epsilon_{\rm ff}$ versus simulation time for a GMC at fixed surface density thresholds of $\Sigma_{\rm gas} = 100$ (blue), $500$ (green), and $1000\, \rm M_\odot\,\mathrm{pc}^{-2}$ (black). The low-density threshold shows a delayed rise in the first 2–3 Myr, while higher thresholds increase earlier. By $\sim$5 Myr, the curves become increasingly similar and evolve in parallel, then show episodic variability with comparable peaks, reaching efficiencies of $\sim$5\%. This demonstrates that efficiency is set by the cloud’s evolution rather than simply the amount of gas enclosed.}
    \label{fig:time_dependence_plot}
\end{figure}

We find that the variation of $\epsilon_{\rm ff}$ across $\Sigma_{\rm gas}$ is most pronounced at early times. 
In the first $3$–$5\,\mathrm{Myr}$ (purple lines), $\epsilon_{\rm ff}$ is very low ($\approx 0.3$–$0.5\%$) at low $\Sigma_{\rm gas}$ and rises sharply toward higher values, producing a brief plateau by $\sim 5\,\mathrm{Myr}$ before decreasing at the highest densities. 
Between $6$–$7\,\mathrm{Myr}$ (orange/red lines), the relation flattens across most $\Sigma_{\rm gas}$ and shows a mild downward trend in denser regions. By late times ($8$–$10\,\mathrm{Myr}$; yellow lines), the relation becomes more stochastic, $\epsilon_{\rm ff}$ remains steady across most $\Sigma_{\rm gas}$ but exhibits local spikes in efficiencies or a more pronounced decline at the highest surface densities as feedback disperses the cloud. 

\subsection{Time Dependence of Star Formation Efficiency}

Figure \ref{fig:time_dependence_plot} shows the temporal evolution of $\epsilon_{\rm ff}$ at three fixed $\log\Sigma_{\rm gas}$ thresholds: $100$, $500$, and $1000\, \rm M_{\odot}\, pc^{-2}$. All three curves share a similar overall evolution. The low-density threshold ($\Sigma_{\rm gas}=100 \, \rm M_{\odot}\, pc^{-2}$; blue line) shows distinct behavior early on, with a delayed rise by nearly an order of magnitude over the first $2$–$4\, \rm Myr$. In contrast, the higher thresholds ($\Sigma_{\rm gas}=500,\, 1000 \, \rm M_{\odot}\, pc^{-2}$; green and black) follow consistent early-time trends with higher initial efficiencies. At intermediate times ($\rm t \sim 4$–$8\, \rm Myr$), $\epsilon_{\rm ff}$ stabilizes and the curves become increasingly similar, evolving in parallel, exhibiting episodic variability with comparable peaks and dips, reaching maximum efficiencies of $\approx 5\%$.

\subsection{Effect of Resolution}

\begin{figure}[htbp] 
    \centering
    \hspace*{-1.1cm}
    \includegraphics[width=0.44\textwidth, trim=01 0 0 0, clip]{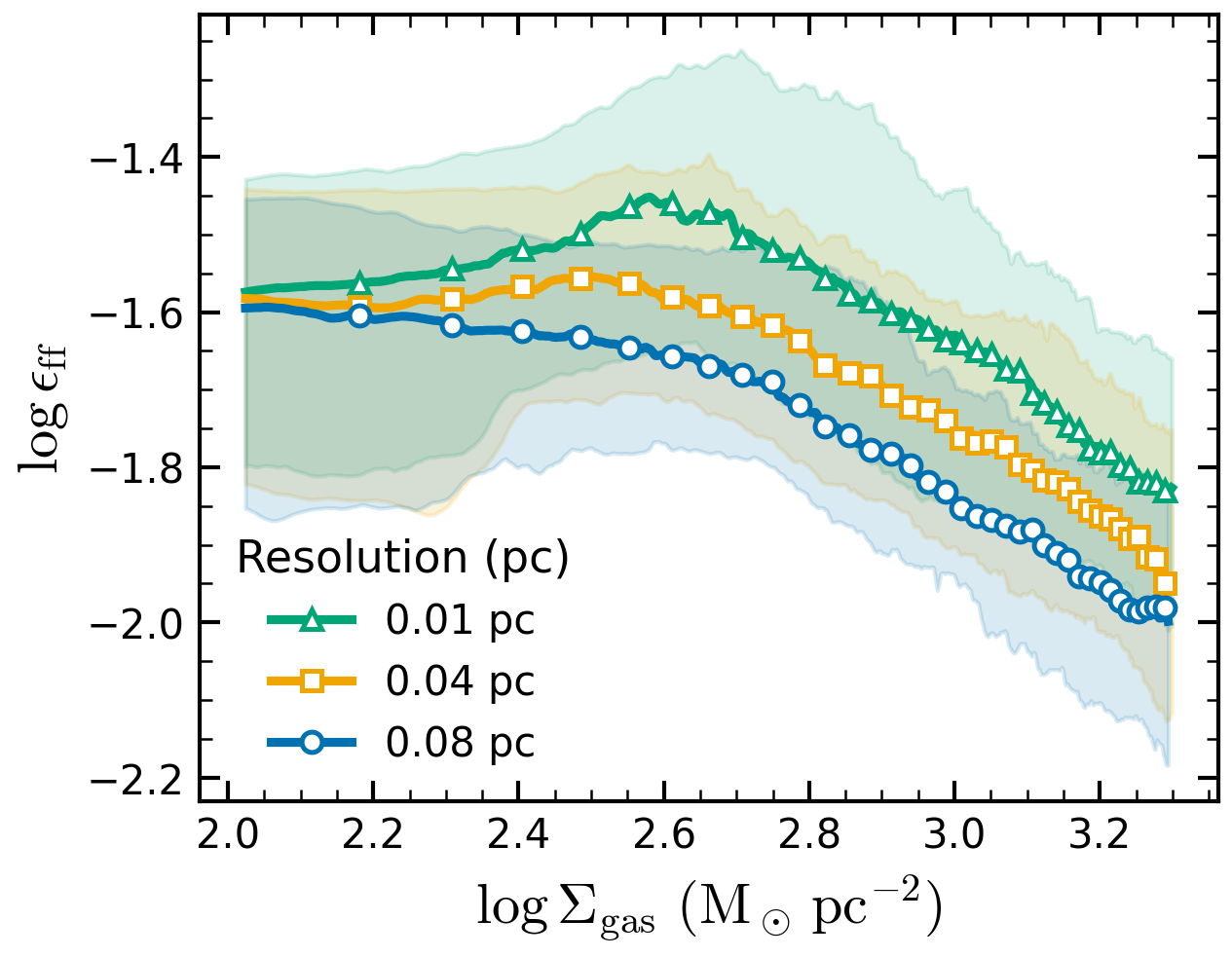}
    \caption{$\epsilon_{\rm ff}$ versus $\Sigma_{\rm gas}$ at three spatial resolutions: $0.01\, \rm pc$ (green, our simulation's native resolution), $0.04\, \rm pc$ (orange), and $0.08\, \rm  pc$ (blue, FIR Herschel/SPIRE Surveys). The curves agree at very low $\Sigma_{\rm gas}$ but start to diverge at early densities (log $\Sigma_{\rm gas} \approx 2.2$), where coarser resolutions smooth over compact gas around YSOs through beam dilution, lengthening free-fall times and lowering inferred efficiencies.}
    \label{Different_resolutions_plot}
\end{figure}

To determine how spatial resolution affects our $\epsilon_{\rm ff}$ measurements, we analyze our synthetic observations at three physical resolutions: $0.01 \, \mathrm{pc}$ (STARFORGE default), $0.04 \, \mathrm{pc}$, and $0.08\, \mathrm{pc}$, with the latter chosen to match the effective resolution of Herschel/SPIRE observations of nearby GMCs \citep{SPIRES2010, pokhrel2020}.
Figure~\ref{Different_resolutions_plot} shows $\log\epsilon_{\rm ff}$  versus $\log\Sigma_\text{gas}$ for the three resolutions. We find that decreasing spatial resolution consistently reduces the inferred $\epsilon_{\rm ff}$. All three resolution values exhibit consistent trends at very low surface densities $(\rm log\, \Sigma_\text{gas} < 2.2 \, \rm M_{\odot}\,  \rm pc^{-2})$, with values centering around $\log\, \epsilon_{\rm ff} \sim -1.6$. However, the different resolutions diverge significantly at increasing values of $\Sigma_\text{gas}$, with a notable convergence of resolutions $0.04$ and $0.08$ pc at higher $\Sigma_\text{gas}$.

\subsection{AGN Injection}

\begin{figure}[b]
    \centering
    \hspace*{-1.1cm}
    \includegraphics[width=0.47\textwidth, trim=09 0 0 0, clip]{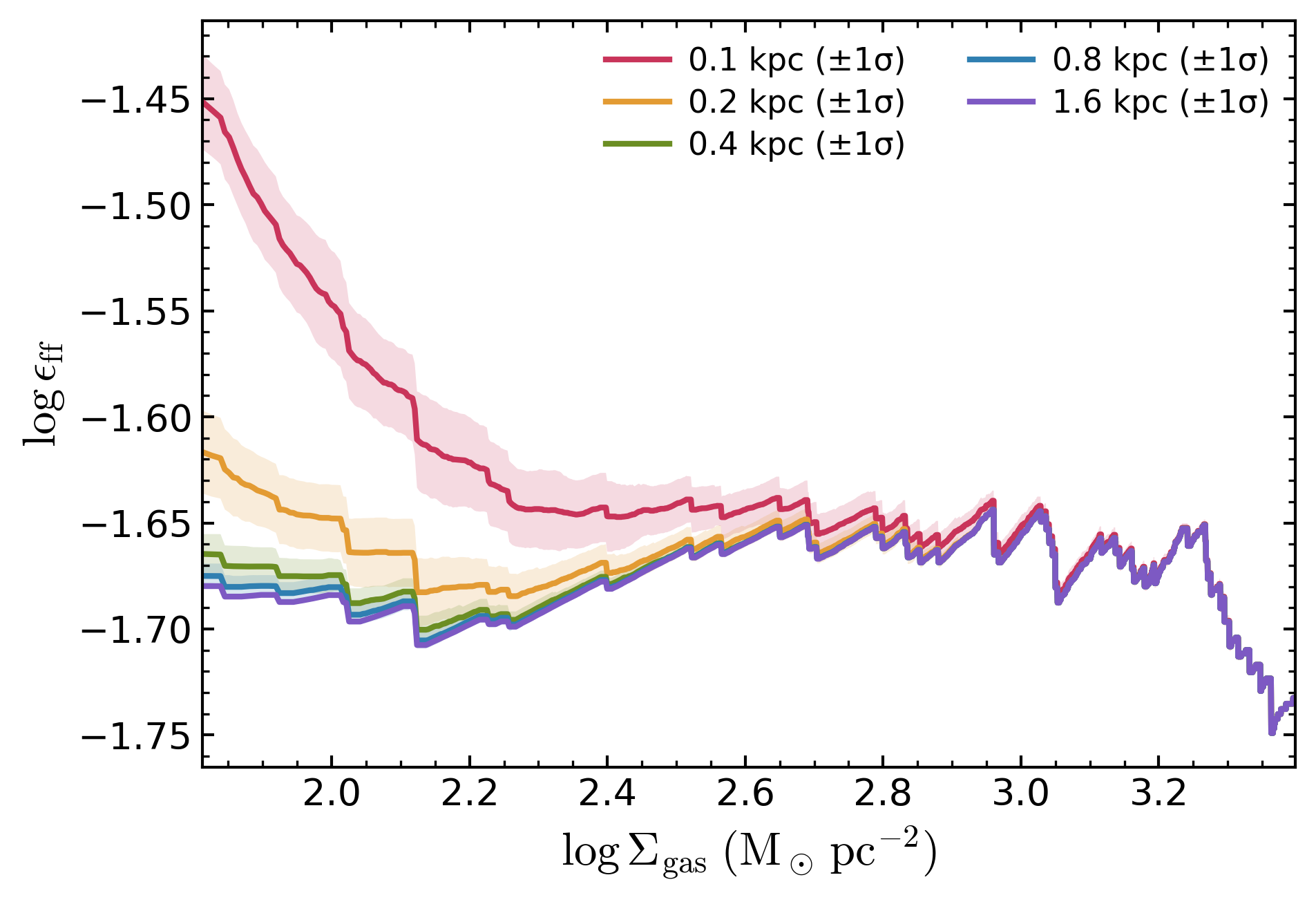}
    \caption{$\epsilon_{\rm ff}$ versus $\Sigma_{\rm gas}$ for clouds placed at distances of 0.1–1.6 kpc, with AGN injected at a surface density of $9~{\rm deg}^{-2}$ \citep{pokhrel2020}. At 0.1–0.4 kpc AGN contaminants elevate $\epsilon_{\rm ff}$ at $\log \Sigma_{\rm gas} \lesssim 2.5 \, \rm M_{\odot}\,  \rm pc^{-2}$. The bias weakens with distance and is negligible by 1 kpc. The snapshot shown corresponds to $t \approx 6$ Myr.}
 \label{ref:AGN}
\end{figure}

To explore how AGN contamination affects the inferred $\epsilon_{\rm ff}$, we inject synthetic AGN into our YSO catalogs using the method described in Section~\ref{sec:Synthetic_Obs}. This simulates the observational scenario where background AGN are misidentified as YSOs within a GMC. Figure~\ref{ref:AGN} shows $\epsilon_{\rm ff}$ versus $\Sigma_{\rm gas}$ for different assumed cloud distances ($0.1$–$1.6\, \rm kpc$) from a sample simulation snapshot at $\rm t \approx 6\,\rm Myr$, along with a statistical $1\sigma$ spread computed over 24 randomized AGN-injection realizations per distance.

The largest spread is seen at $0.1–0.4\, \rm kpc$, with $\epsilon_{\rm ff}$ offset by up to $\sim$ 0.1 dex at low surface densities ($\log \Sigma_{\rm gas} \lesssim 2.5\, \rm M_{\odot}\,  \rm pc^{-2}$). At larger distances, the spread decreases, and by $1\, \rm kpc$ the curves show only minor differences, with values remaining similar across the surface density range.

\setlength{\tabcolsep}{30pt} 
\begin{deluxetable*}{ccc}
\tablecaption{\add{Alternate expressions of $\epsilon_{\rm ff} = \mathrm{SFR}/\mathrm{FCR}$ with the variant's definition of the star formation rate (SFR) and the free-fall collapse rate (FCR). \label{tab:epsff_variants}}}

\tablewidth{\textwidth} 

\tablehead{
\colhead{Variant} & \colhead{Star Formation Rate (SFR)} & \colhead{Free-fall Collapse Rate (FCR)}
}

\startdata
\textbf{Observational} & $\mathrm{SFR}_{\mathrm{Obs}} = \frac{N_{\rm YSO}\,\langle M_{\rm YSO}\rangle}{\langle t_{\rm YSO}\rangle}$ & $\mathrm{FCR}_{\mathrm{Obs}} = \frac{M_{\rm gas}}{t_{\rm ff}^{\rm sph}}$ \\
\textbf{Simulation}    & $\mathrm{SFR}_{\mathrm{Sim}} = \big\langle \dot{M}_\star \big\rangle_{\Delta t}$ & $\mathrm{FCR}_{\mathrm{Sim}} = \sum_i \frac{m_{{\rm gas},i}}{t_{{\rm ff},i}}$ \\
\textbf{Mixed A}       & $\mathrm{SFR}_{\mathrm{Sim}}$ & $\mathrm{FCR}_{\mathrm{Obs}}$ \\
\textbf{Mixed B}       & $\mathrm{SFR}_{\mathrm{Obs}}$ & $\mathrm{FCR}_{\mathrm{Sim}}$ \\
\enddata

\tablecomments{\add{$\dot{M}_\star$ is the total stellar accretion rate averaged over a time window $\Delta t$. 
$t_{\rm ff}^{\rm sph}$ is the free-fall time assuming a uniform sphere. 
$\mathrm{FCR}_{\mathrm{Sim}}$ sums the gas particle masses divided by their respective free-fall times for all particles within the contour. 
$\mathrm{SFR}_{\mathrm{Obs}}$ corresponds to Class $0$ YSO-counting assumptions with a fixed mean stellar mass 
($\langle M_{\rm YSO}\rangle \approx 0.5\,M_\odot$) and a mean lifetime 
($\langle t_{\rm YSO}\rangle \approx 0.5\,\mathrm{Myr}$).}}
\end{deluxetable*}

\section{Discussion}\label{sec:discussion}

In this work, we generated and analyzed mock FIR observations of STARFORGE’s fiducial star cluster formation simulation suite to reproduce the observationally inferred star formation efficiencies ($\epsilon_{\rm ff}$) of star-forming environments, following the methods of \citet{Pokhrel2021}. This approach allowed us to bridge the gap between computational estimates derived directly from our simulated outputs and those inferred from protostellar counts and FIR-based mass estimates. Our results demonstrate that $\epsilon_{\rm ff}$ varies over a cloud's lifetime and is regulated by stellar feedback.

Numerous simulations have demonstrated that stellar feedback is a key mechanism in regulating star formation in low-mass and massive star-forming clouds. However, due to resolution constraints, star cluster formation simulations are often unable to resolve individual stars on massive GMC scales ($\gtrsim 10^4~\rm{M_\odot}$) and therefore only include feedback from massive stars \citep[e.g.,]{Suin2024} or resolve individual star formation in collapsing low-mass star-forming clouds or regions ($\lesssim 10^3~\rm{M_\odot}$) that do not form massive stars \citep[e.g.,][]{offner2009, Federath2015, cunningham2018, Appel+2023}. 

A key advantage of the \add{simulations presented here is that STARFORGE properly models the formation of stellar systems} across the stellar mass spectrum (down to $0.1~\rm{M_\odot}$) on massive GMC scales ($\sim 10-100$~pc) and includes all feedback mechanisms inherent to star formation. This has allowed us to utilize the methods of \citet{Pokhrel2021} to infer star formation rates from protostellar (Class 0) counts and estimated lifetimes ($\sim 0.5$~Myr) motivated by prior observational studies \citep{Evans2009, Dunham2014}. 

\subsection{Simulation vs. Observational Interpretations of $\epsilon_{\rm ff}$} \label{theory_obser_discussion}

In addition to stellar feedback, methodological choices are likely an important contributor to the discrepancy between simulation and observationally determined values of $\epsilon_{\rm ff}$. To examine this, we compare definitions directly (see Table~\ref{tab:epsff_variants}), setting observational and simulation formulations of star formation rates and free-fall collapse rates (FCRs) side by side, and introduce hybrid expressions to test where the two perspectives agree or disagree. Our mixed approaches (Table~\ref{tab:epsff_variants}, rows 3 and 4) act as diagnostics that isolate these effects. For example, Mixed Method A combines observational star formation rates, which assumes a common protostar mass estimate $(0.5~\rm M_\odot)$ and Class 0 lifetime $(0.5~\rm Myr)$, with simulation FCRs for each particle within the contour.

\begin{figure*}[!t]
    \hspace*{-0.7cm}
    \centering
    \includegraphics[width=0.79\textwidth]{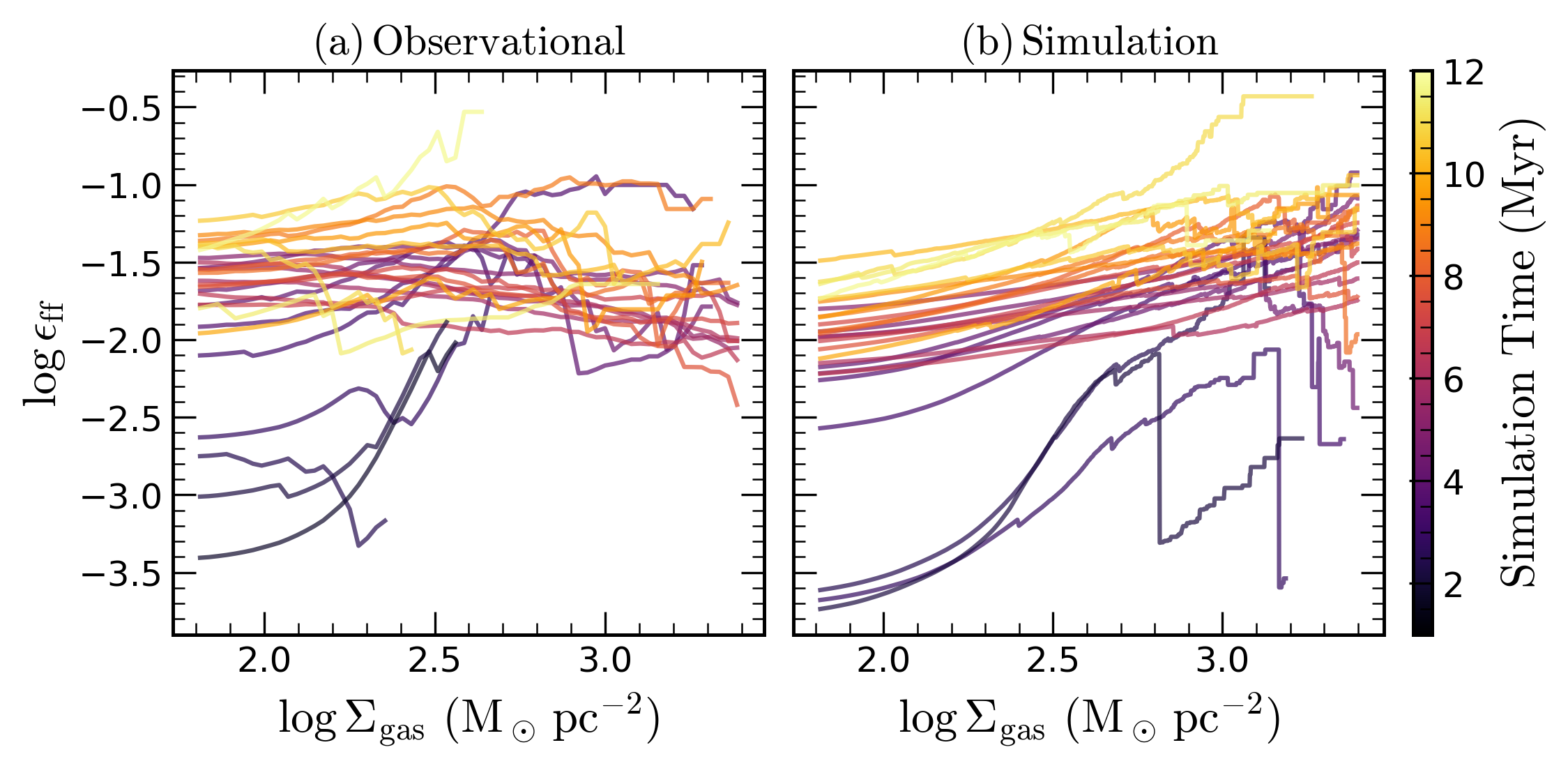}
    \caption{$\epsilon_{\rm ff}$ versus $\Sigma_{\rm gas}$ for the fiducial GMC over 3–10 Myr. (a) The observational approach used in this study. (b) Simulation based on stellar accretion histories and multi–free-fall times. Curves are color-coded by cloud age, with purple denoting early times and yellow denoting late times. The observational approach shows greater scatter at most $\Sigma_{\rm gas}$, whereas the simulation approach increases more systematically with increasing $\Sigma_{\rm gas}$. This exemplifies how methodological choices shape $\epsilon_{\rm ff}$ calculations.}
    \label{fig:comparison_plot}
\end{figure*}

Figure~\ref{fig:comparison_plot} compares the different simulation and observational estimates for $\epsilon_{\rm ff}$ described in Table~\ref{tab:epsff_variants}. Both the observational estimates from our mock observations and ``ground truth" values inferred from the simulation exhibit scatter across the cloud’s evolution.

However, the simulation values capture the systematic rise in $\epsilon_{\rm ff}$ at high densities that observations miss. At first glance, this bias showcases the reliance on fixed average protostellar masses and lifetimes, consequently adopting a constant SFR. \add{This bias was similarly identified in prior numerical studies by \citet{Glover_Clark_2014}, though they primarily focused on later-stage YSO classes I through III and emphasized interstellar radiation field (ISRF) and dust shielding rather than self-consistent cluster formation with stellar feedback. They assumed a much longer average YSO age of $\sim 2$ Myr but adopted the same average mass as our Class 0 YSOs.}

This average corresponds to the peak of the system IMF and a Class 0 average lifetime of $0.5\, {\rm Myr}$, yet massive protostars violate both assumptions through rapid accretion and shortened protostellar lifetimes due to fast Kelvin Helmholtz contraction timescales \citep{2007Motte, Rosen2016a, Rosen2022a}. As a result, the observational instantaneous star formation rate estimates are systematically underestimated in dense regions where massive stars form. \add{Similarly, at lower density regions, we have few stars being born, consequently breaking down the statistical sampling of the system IMF. In this regime, $0.5\, \rm M_\odot$ may be too large an average as there are simply not enough stars to populate the distribution, and the justification of a fixed mean stellar mass becomes increasingly uncertain.}

\begin{figure*}[!t]
    \hspace*{-1.2cm}
  \centering
  \includegraphics[width=0.65\textwidth]{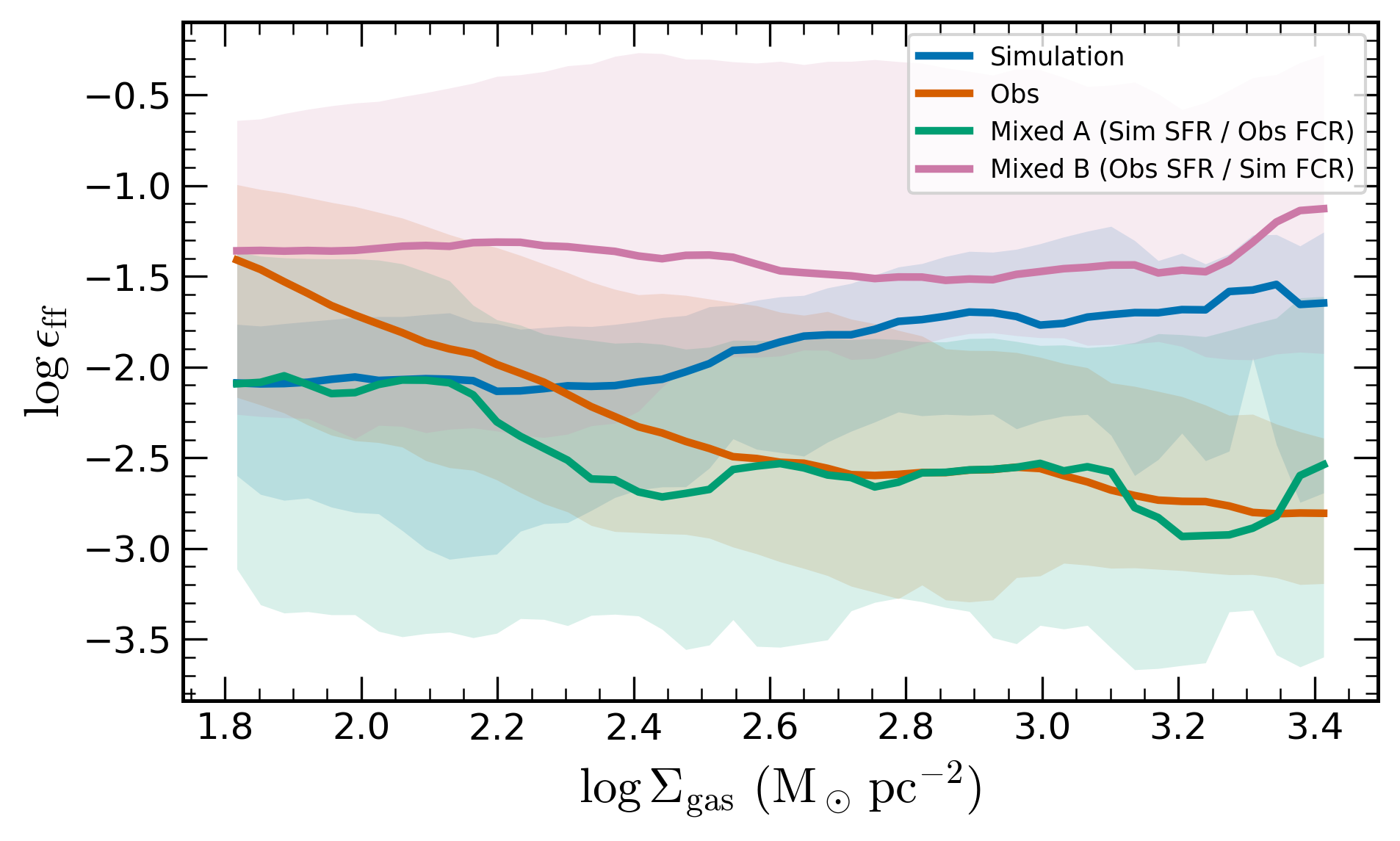}
  \caption{$\epsilon_{\rm ff}$ versus $\Sigma_{\rm gas}$ for different methodological definitions define in Table \ref{tab:epsff_variants}. Shaded regions show the $1\sigma$ scatter. The observational form (red) yields higher values at low $\Sigma_{\rm gas}$, due to biases in the SFR  while still reasonably capturing the less dense gas, but falls below the simulation values (blue) at higher densities. At high $\Sigma_{\rm gas}$, geometrical assumptions smooth dense structure, that leads to longer inferred free-fall times and increased enclosed gas mass, which suppresses $\epsilon_{\rm ff}$. The two approaches cross near $\log \Sigma_{\rm gas} \approx 2.3 \rm M_{\odot} {\rm pc}^{-2}$, marking a transition threshold before diverging with increasing $\Sigma_{\rm gas}$. Hybrid expressions (green, purple) combine numerators and denominators from each method, illustrating how assumptions of gas structure and star formation tracers drive the offsets}
  \label{fig:master_plot}
\end{figure*}

However, Figure~\ref{fig:master_plot} demonstrates that SFR assumptions do not entirely drive the methodological differences. Our hybrid formulations instead highlight the geometrical treatment of the cloud and free-fall collapse as the primary driver. Mixed B and the ground truth values adequately capture the systematic rise with $\log \Sigma_{\rm gas}$, while Mixed A and the observational values do not. This points towards the geometrical assumptions of molecular clouds that inaccurately portray the morphology, effectively smoothing over dense structure and redistributing mass, leading to an overestimation of free-fall times and an increase in enclosed gas at increasing $\log \Sigma_{\rm gas}$ , which ultimately suppresses $\epsilon_{\rm ff}$. This is consistent with \citet{Zipeng2022}, who show that spherical geometry assumptions systematically overestimate free-fall times. Additionally, Figure~\ref{fig:master_plot} highlights a threshold around $\log \Sigma_{\rm gas} \approx 2.3, \rm M_{\odot}, {\rm pc}^{-2}$ that marks where these effects begin to dominate, but we emphasize that this threshold arises from methodological and resolution choices rather than reflecting a universal physical boundary. Below this threshold, observational methods yield higher $\epsilon_{\rm ff}$ than simulation estimates, likely due to an overestimation of the SFR while still reasonably capturing the less dense gas. Above the threshold, the trend reverses, reflecting an underestimation of the SFR and a poorer representation of the enclosed gas due to the smoothing of dense structure. This surface density may correspond to the regime where massive star formation becomes important, at which point both observational assumptions begin to break down.

\add{Beyond the assumptions of mass and lifetime, one important caveat that warrants discussion is that we do not attempt to identify Class 0 YSOs from our synthetic FIR maps. Instead, we adopt a hybrid approach that identifies Class 0 objects through formation times recorded by the simulation and associates each protostellar sink particle with the local $\Sigma_{\rm gas}$ measured from our synthetic maps. Creating observational column densities from simulations is well established \citet{Shetty2011a}, but YSO identification is inherently non-trivial and complex, involving SED fitting and classification algorithms that are sensitive to contamination and projection effects \citep{Evans2009}. Our work primarily focuses on reproducing observational measurements of $\epsilon_{\rm ff} \sim 1$–$3\%$ and determining if the discrepancy arises from physical processes not included in previous studies. This approach allows us to minimize missing YSO counts and misclassifications to accurately depict SFRs. Consequently, we can isolate potential biases in the gas dynamics that affect protostellar birth without compounding the systematic uncertainties in a full observational methodology. Implementing a fully observational YSO identification pipeline would likely increase scatter and shift the SFR within specific contours, but it would not remove the key trends we encounter here: the strong time dependence and the resolution dependence of the measured $\epsilon_{\rm ff}$}. In essence, stellar feedback regulates the efficiency of star formation, while observational methodology and resolution determine how it's inferred.

\subsection{Time-Dependent Evolution}
By taking into account the time evolution of star-forming GMCs, our results reveal that $\epsilon_{\rm ff}$ is not a static quantity. Following the evolution of $\epsilon_{\rm ff}$ as a function of both $\Sigma_{\rm gas}$ and time, we show that $\epsilon_{\rm ff}$ evolves dynamically throughout the cloud’s life cycle. As shown in Figure~\ref{fig:time_dependence_plot}, a single GMC exhibits significant scatter over its $\sim$$10\,\mathrm{Myr}$ lifetime across different $\Sigma_{\rm gas}$ ranges, with episodic variations that challenge the interpretation of $\epsilon_{\rm ff}$ as a universal relation between gas supply and star formation. 

Our results suggest that the scatter reported by \citet{Pokhrel2021} can be explained by evolutionary stage rather than intrinsic physical differences between GMCs. The variations within a single simulated cloud mirror those seen across surveys, reflecting that observations capture clouds at different points along a shared evolutionary track. Bursts of high efficiency are linked to massive star formation and the feedback it injects, underscoring their role in driving the evolution of $\epsilon_{\rm ff}$ which eventually reduces $\epsilon_{\rm ff}$ at late stages of the cloud's evolution before cloud dispersal occurs. As a result, attempts to link $\epsilon_{\rm ff}$ to specific mechanisms may be constrained unless the cloud’s evolutionary state is explicitly considered. Recognizing this time dependence is therefore essential for connecting theoretical predictions with observational measurements of $\epsilon_{\rm ff}$.

\subsection{Observational Biases}

Our mock observations reveal that FIR survey systematics strongly bias $\epsilon_{\rm ff}$, with spatial resolution being the dominant source, and AGN contamination a secondary but noticeable bias that depends on cloud distance. As shown in Figure~\ref{Different_resolutions_plot}, lowering the simulation spatial resolution systematically reduces the measured $\epsilon_{\rm ff}$ at most surface densities. Coarse resolution smooths compact structures around protostars, lowers peak surface densities, and shrinks dense contours, which redistributes YSOs and lowers observationally inferred estimates of $\epsilon_{\rm ff}$. Future FIR instruments with sub-$0.1\,\rm pc$ resolution could alleviate this by resolving the small-scale structure around protostars, but current instruments lack this capability. Until then, observational studies must account for resolution smoothing or risk attributing instrumental limits to underlying physical mechanisms.  

Beyond resolution constraints, AGN contamination introduces a smaller bias in our analysis. Shown in Figure \ref{ref:AGN}, its influence is confined mainly to nearby clouds and lower-density regimes, where background sources can blend into YSO catalogs through their FIR emission. While this produces a modest upward shift in $\epsilon_{\rm ff}$ at the low end of $\Sigma_{\rm gas}$, the effect weakens with distance and becomes negligible compared to the resolution-driven biases. These results imply that resolution is the dominant observational bias in FIR studies of $\epsilon_{\rm ff}$, while AGN contamination plays only a secondary role. Accounting for both is critical for connecting observational measurements to theoretical predictions. Observers should remain mindful of these biases when interpreting efficiencies, and theorists applying observational techniques to simulations must do the same to ensure consistent comparisons.

\subsection{STARFORGE Limitations}

STARFORGE is the most physically complete simulation suite of star formation to date, but it still relies on assumptions and idealizations. The GMC is initialized as a uniform spherical cloud with a fixed mass, radius, metallicity, and uniform magnetic field strength, whose initial velocity distribution is seeded by turbulent motions. These idealized initial conditions does not capture the complex and poorly constrained assembly histories of real molecular clouds. As the simulation evolves, turbulence is allowed to freely decay rather than being continuously driven, unlike in the Milky Way where large-scale flows, feedback, and shear replenish it. 

Additional physical processes are also treated simplistically, including the treatment of cosmic ray heating, grain-surface chemistry dust physics,
and radiative transfer that employs an M1 solver and reduced speed of light approximation, which can underestimate radiation pressure in complex geometries. These choices keep the problem computationally feasible but can alter gas temperatures, abundances, and feedback strength, all of which feed into star formation rates and $\epsilon_{\rm ff}$. These limitations underscore the difficulty of modeling star formation in full detail but also help frame the scope of this study. STARFORGE provides a platform to test how feedback regulates $\epsilon_{\rm ff}$, with efficiencies that reach observed ranges without extreme initial conditions and evolve naturally over a cloud’s lifetime. 

\section{Conclusion}\label{sec:conclusion}

In this work, we generated synthetic FIR observations of a high-fidelity RMHD STARFORGE simulation, which models massive star cluster formation via the gravitational collapse of a massive, turbulent GMC that self-consistently models star formation and stellar feedback in the form of radiation, collimated outflows, and stellar winds. This approach has allowed us to directly apply observational techniques employed in FIR GMC star formation surveys, which rely on counting individual protostars,  to determine the evolution of the observationally-inferred $\epsilon_{\rm ff}$ as a function of cloud evolution from the onset of star formation to cloud dispersal due to feedback. We find that stellar feedback regulates $\epsilon_{\rm ff}$, producing evolutionary variations in $\epsilon_{\rm ff}$ that are dependent on local cloud conditions and prior star formation. By comparing our observationally-inferred estimates of $\epsilon_{\rm ff}$ directly to the true star formation rates taken directly from the simulated outputs (i.e., the ``ground truth") we were able to determine
limitations that can arise from inferring $\epsilon_{\rm ff}$ from FIR star formation studies that often employ simplified assumptions, including a fixed protostellar mass and lifetime to estimate the SFR and geometrical assumptions to quantify the morphology fo the cloud . We reach the following conclusions.

\begin{itemize}  
    \item $\epsilon_{\rm ff}$ varies strongly with surface density and cloud evolutionary state. In our high-resolution GMC simulation, efficiencies span over an order of magnitude (0.05-10\%) across $\sim$10~Myr, naturally reproducing the low $\epsilon_{\rm ff} \approx 1-3\%$ values inferred from observations. This natural time dependence helps explain the scatter in surveys, demonstrating that $\epsilon_{\rm ff} \approx 1-3\%$ is regulated by stellar feedback and highly variable throughout the lifetimes of star-forming GMCs.  

   \item We find that our computational estimates of $\epsilon_{\rm ff}$ differ systematically from observationally-inferred calculations. Below $\log \Sigma_{\rm gas} \approx 2.3~{\rm M_{\odot},pc^{-2}}$, observations tend to overestimate $\epsilon_{\rm ff}$, likely due fixed assumption in the SFR formulation. Above this threshold, the trend reverses; smoothing of dense structure leads to longer inferred free-fall times and larger enclosed gas, suppressing $\epsilon_{\rm ff}$, while fixed protostellar masses and lifetimes further contribute to underestimation of the SFR.

    \item Limited resolution suppresses estimated $\epsilon_{\rm ff}$ in almost all regions, while AGN contamination biases measurements in less dense regions of the cloud. Coarse resolution (0.08~pc vs. 0.01~pc) lowers efficiencies by up to 0.3 dex throughout $\Sigma_{\rm gas}$,  while AGN contamination produces smaller ($<0.1$ dex) effects in nearby clouds at low $\Sigma_{\rm gas}$. 
\end{itemize}

This work demonstrates that synthetic observations of high-fidelity star cluster formation simulations that resolve stellar systems and stellar feedback on GMC scales provide a comprehensive framework that bridges computational and observational estimates of star formation metrics. Our pipeline reproduces the observed $\epsilon_{\rm ff}$ in Milky Way star-forming clouds while allowing us to follow this quantity throughout a GMC's lifetime. Our results suggest that the large scatter in observationally-inferred estimates of $\epsilon_{\rm ff}$ may not reflect intrinsic differences between clouds, but instead arises naturally from their evolutionary stage and observational biases.

\begin{acknowledgments}
A.E. acknowledges support through the Carnegie Astrophysics Summer Student Internship (CASSI) and San Diego State University's Undergraduate Research Program (SURP). A.E. thanks Anna Rosen (SDSU) and Michael Grudi\'{c} (Flatiron Institute) for insightful conversations and guidance that shaped this work, and the anonymous reviewer for a thoughtful and constructive report that improved the clarity and quality of this manuscript. The STARFORGE simulations used in this work were produced on the Frontera supercomputer at the Texas Advanced Computing Center (TACC), supported by the National Science Foundation (NSF) under grant OAC-1818253, and analyzed through the computational resources provided by the Flatiron Institute, supported by the Simons Foundation. 
\software{STARFORGE \citep{starforgemethods}, Gizmo \citep{hopkinsgizmo2017}}
\end{acknowledgments}


\end{document}